\documentclass[twocolumn,showpacs,preprintnumbers,amsmath,amssymb]{revtex4}
\usepackage{amsfonts}
\usepackage{amsmath}
\usepackage{graphicx}
\usepackage{dcolumn}
\usepackage{bm}
\usepackage{overpic}
\usepackage{booktabs}

\begin{document}
\title{Coarse Graining for Synchronization in Directed Networks}
\author{An Zeng, Linyuan L\"u}\email{linyuan.lue@unifr.ch}
 \affiliation{Department of Physics, University of Fribourg, Chemin du Mus\'{e}e 3, CH-1700 Fribourg, Switzerland}

\date{\today}

\begin{abstract}
Coarse graining model is a promising way to analyze and visualize
large-scale networks. The coarse-grained networks are required to
preserve statistical properties as well as the dynamic
behaviors of the initial networks. Some methods have been proposed
and found effective in undirected networks, while the study on
coarse graining directed networks lacks of consideration. In this
paper, we proposed a Path-based Coarse Graining (PCG) method to
coarse grain the directed networks. Performing the linear stability
analysis of synchronization and numerical simulation of the Kuramoto
model on four kinds of directed networks, including tree
networks and variants of Barab\'{a}si-Albert networks,
Watts-Strogatz networks and Erd\"{o}s-R\'{e}nyi networks, we find
our method can effectively preserve the network synchronizability.


\end{abstract}



\maketitle

\section{Introduction.}
Complex networks have become a key approach to understanding many
social, biological, chemical, physical and information systems,
where nodes represent individuals and links denote the relations or
interactions between nodes. In this sense, to study the dynamics of
such systems is actually to investigate the dynamical behaviors on
the networks. In particular, the network synchronization as an
important emerging phenomenon of a population of dynamically
interacting units in various fields of science has attracted much
attention~\cite{PR93,Ren2010,PRL014101,PRE067105,PRL0341012,Zhao2006,Wu2006,PRL034101,PRE016116,PRL218701,PRL138701,ZhaoEPJB2006}.
Most works focused on studying the relation between network topology
and the
synchronization~\cite{PRL014101,PRE067105,PRL0341012,Zhao2006,Wu2006},
enhancing the synchronizability by designing the weighting
strategies~\cite{PRL034101,PRE016116,PRL218701,PRL138701,ZhaoEPJB2006}.
Moveover, some efforts have been made to study the synchronization
in directed
networks~\cite{PRE065106,PhysicaD224,EPL48002,NJP043030,PNAS10342}.
It has been pointed out that the optimal structure for
synchronizability is a directed tree~\cite{PRE065106,PhysicaD224} and
the convergence time is strongly related to the depth of the
tree~\cite{EPL48002,NJP043030}. Most of the experiments on
investigating the dynamic behaviors are implemented on small-size
networks. However, when the networks contains very large number of
nodes, it becomes sometimes impossible to model the dynamic process.
For example, to investigate the synchronization, extrapolating the
coupled differential equations model of a single node to this large
system is too complicated to be carried out.

A promising way to address this problem is to coarse grain the
network, namely to reduce the network complexity by means of mapping
the large network into a smaller one. The coarse graining techniques
have been successfully applied to model large genetic networks~\cite{Bornholdt2005} and extract the slowest motions in protein
networks~\cite{Kurkcuoglua2004}. Essentially, the coarse graining
process is very similar to the problem of cluster finding or
community detection in networks (see
Ref.~\cite{Girvan2002,Reichardt2004,Danon2005,Fortunato2010} for some
popular methods). The coarse-grained network is obtained by merging
the nodes in the same cluster or community. However, the coarse
graining model is far beyond the clustering techniques, since it
requires the coarse-grained networks keep the same topological
properties or dynamic behaviors as the initial networks, such as
preserving the degree distribution, cluster coefficient,
assortativity correlation~\cite{PRL168701}, the properties of random
walk on the network~\cite{PRL038701}, synchronization~\cite{PRL174104} and critical
phenomena~\cite{PRE011107}. Most
of the former works on coarse graining consider undirected networks.
However, in many real systems, the interactions between individuals
are not reciprocal. For example, the food web, gene regulation
system, metabolic system and neural system are usually represented
by directed networks where the nodes are affected by their upstream
nodes. In directed networks, of course we can ignore link directions
and apply methods developed for undirected
networks, but this approach discarding potentially useful
information contained in the link directions may lead to
dramatically change of the key organizational features when coarse
graining the networks~\cite{PRE016103}. In addition, some prominent
methods may confront problems when applied to directed networks.
Among all these existing coarse graining methods for undirected networks, the spectral coarse graining (SCG) method is a very
general method which can be applied in many dynamic processes such as synchronization, random walk and epidemic spread~\cite{PRL174104,PRL038701}. In order to preserve a typical eigenvalue, the SCG method coarse grains the nodes with similar elements in the corresponding eigenvectors. For different dynamic processes, different
eigenvalues should be considered. Therefore the choice of the eigenvectors is indeed problem dependent. As the synchronizability is mainly related to the largest and smallest nonzero eigenvalue,
 the SCG method for synchronization takes the $p_{2}$ and $p_{N}$ into consideration ($p_{2}$ and $p_{N}$ are respectively the eigenvectors for the smallest nonzero and the largest eigenvalue). However, this method may not provide good
performance in directed networks since the eigenvector elements cannot successfully characterize the nodes' dynamic role. For instance, the nodes in different layers may have exactly the same eigenvector elements in directed acyclic networks while the nodes with exactly the same topology may have totally different eigenvector elements in directed networks with cycles. In a word, to design an effective coarse graining method for directed networks is still challenging.

In this paper, we propose a Path-based Coarse Graining (PCG)
method to coarse grain directed networks for synchronization. The
basic idea is that the nodes who obtain the same impacts from other
nodes are similar to each other, and thus can be merged. The impacts
that one node receives from other nodes are calculated via tracing
the origin of the source in the directed networks (i.e., along the
opposite direction of links). It has been pointed out that the dynamical correlation can be predicted from such topological similarity ~\cite{PRE026116}. Therefore, coarse graining in this way will most naturally merge the nodes according to their functional performance and likely preserve the dynamical properties. The linear stability analysis of
synchronization and numerical simulation of the Kuramoto model on
four kinds of directed networks, including tree networks and
variants of Barab\'{a}si-Albert networks, Watts-Strogatz networks
and Erd\"{o}s-R\'{e}nyi networks, show that our method can
effectively preserve the synchronizability of the initial directed
networks. Additionally, we find the far sources play more important
roles when identifying the nodes' roles in directed networks with
obvious hierarchy structure, while the near sources are more
important in the directed networks with many loops.

\section{Path-based Coarse Graining (PCG) Method}
\subsection{Definition of node's impact-vector}
Many structural-based similarity indices have been proposed to
quantify the nodes' similarity~\cite{EPJB623,PRE046122,EPL58007},
most of which only work for undirected networks. How to define the
nodes' similarity in directed networks is still a challenge. Here we
propose a method via tracing the origin of impacts in directed
network. The basic assumption is that two nodes are
structural-similar if they obtain the same impacts from other nodes,
and thus they are more likely to be merged during the coarse
graining process. Given a directed network $G(V,E)$, where $V$ and
$E$ denote the set of nodes and directed links respectively.
Multiple links and self-connections are not allowed. The impact of
node $x$ on node $y$ is defined by summing over the collection of
directed paths from $x$ to $y$ with exponential weights by length.
The mathematical expression reads
\begin{equation}
f_{x\rightarrow y}=\sum\limits_{l=1}\limits^{l_{max}}
\beta^{l}|path_{x\rightarrow y}^{<l>}|,\label{tracing}
\end{equation}
where $|path_{x\rightarrow y}^{<l>}|$ is the set of all directed
paths with length $l$ starting from node $x$ to node $y$.
Mathematically, $|path_{x\rightarrow y}^{<l>}|=(A^l)_{xy}$, where
$A$ is the adjacency matrix: if $x$ points to $y$ $A_{xy}=1$,
otherwise $A_{xy}=0$. $\beta$ is a free parameter that controls the
weights of the paths. Smaller $\beta$ indicates assigning more
weights on the short paths, and vice versa. Here the paths whose
lengths are not larger than $l_{max}$ are considered. If
$l_{max}=\infty$, namely considering all directed paths from $x$ to
$y$, Eq.~\ref{tracing} is similar to the \emph{Katz} index~\cite{PC39}. However, the significant difference is twofold: On one
hand, the adjacency matrix in Katz index is symmetrical while
asymmetrical in Eq.~\ref{tracing}; On the other hand, the parameter
$\beta$ is usually smaller than unit in Katz index, namely assigning
more weights to the short paths, while in Eq.~\ref{tracing} $\beta$
has no limitation. Since counting all paths between every pair of
nodes is very time-consuming especially in large networks, we here
set $l_{max}$ equal to the length of the longest path among all the
shortest paths between two nodes. Note that when $l_{max}=\infty$
and $\beta$ is smaller than the reciprocal of the largest eigenvalue
of $A$ (i.e., ensure the convergence), the impact matrix $F$ with
element $f_{xy}$ defined in Eq.~\ref{tracing} can be directly
calculated by $F=(I-\beta A)^{-1}-I$.

\subsection{Group partition via k-means clustering}
We assign each node $x$ a $N$ dimensional impact-vector which is
equal to the $x$th column of matrix $F$, namely
$\vec{f}_x=(f_{1x},f_{2x},f_{3x},\cdots,f_{Nx})^T$, where $N=|V|$ is
the number of nodes. Clearly, if two nodes receive the same impacts
from their ancestors (i.e., upstream nodes), they tend to have the same
phase in synchronization, and thus are more likely to be merged
during coarse graining. Suppose we are going to coarse grain a
network containing $N$ nodes to a smaller one with $K$ nodes. We
adopt the k-means clustering method~\cite{MacQueen1967} to partition
the $N$ nodes into $K$ groups. The nodes in the same group will be
merged. The k-means clustering technique aims at minimizing the
within-cluster sum of squares:
\begin{equation}
E=\sum\limits_{i=1}\limits^{K} \sum\limits_{x\in V(i)}
||\vec{f}_x-\vec{c}(i)||^2,
\end{equation}
where $V(i)$ is the set of nodes in cluster $i$ ($i=1,2,\cdots,K$),
and $\vec{c}(i)$ is the centroid of cluster $i$ which is equal to
the mean of points in cluster $i$, namely
\begin{equation}
c_k(i)=\frac{1}{|V(i)|}\sum\limits_{x\in
V(i)}f_{kx}.\label{centroid}
\end{equation}
The detailed steps of k-means clustering are shown as follows: (i)
Choose $K$ vectors as the initial centroid of each cluster. (ii)
Randomly choose a node $x$ from the set $V$. This node will belong
to the cluster $i$ if the distance between its vector $\vec{f}_x$
and the centroid of cluster $i$, namely $\vec{c}(i)$, is the minimum
among all the centroids of $K$ clusters. (iii) Update the centroid
of each cluster according to Eq.~\ref{centroid}. (iv) Repeat steps
(ii) and (iii) until all the centroids cannot be modified. Note that
for a given $K$, clusters will depend heavily on the initial
configuration of the set of centroids, thus making interpretation of
the clusters rather uncertain. Different initialization may lead to
different solutions which may trapped in the local minimum. Clusters
should be, as much as possible, compact, well separated, and
interpretable, possibly with the help of some additional variables,
such as the $F$-statistic. Here we only focus on whether the
clusters are compact, namely the vectors (nodes) within one cluster
are close (similar) enough, while neglect if the clusters are well
separated. Therefore we will finally choose the clustering result
subject to the lowest $E$ among $L$-possible solutions (we set
$L=20$ in this paper).

\subsection{Weighting strategy for the coarse-grained networks}
\begin{figure}
  \center
  \includegraphics[width=8cm]{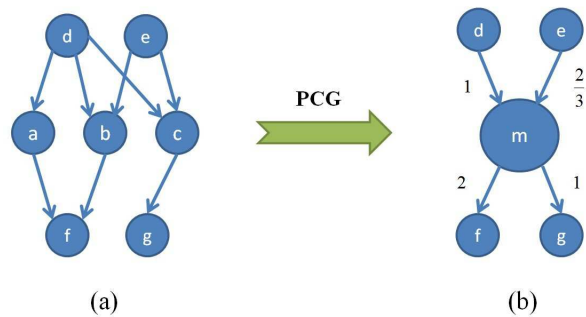}
\caption{(Color online) A simple illustration of how to update the
links' weights in the coarse graining process. (a) shows the initial
network constituted of seven nodes and eight directed links. (b) is
the reduced network after merging the three nodes $a$, $b$ and $c$.
Numbers on the links indicate the new weights of the
links.}\label{fig1}
\end{figure}

Another crucial problem in the process of coarse graining is how to
update the links' weights after merging the nodes so that the
resulting network is truly representative of the initial one. An
effective weighting strategy was proposed by Gfeller \emph{et al.}~\cite{PRL174104}. Here we apply it to directed networks.
Specifically, when we merge the nodes in cluster $i$ to form a new
node labeled by $m_i$, the weights of the merged links update
according to the following principle:
\begin{equation}
\begin{cases}
w_{x\rightarrow m_i}=\frac{\sum\limits_{y\in V(i)}w_{x\rightarrow y}}{|V(i)|}, \quad \texttt{for $m_i$'s in-links}\\
\\
w_{m_i\rightarrow x}=\sum\limits_{y\in V(i)}w_{x\rightarrow y},
\quad \texttt{for $m_i$'s out-links}
\end{cases}\label{weight}
\end{equation}
where $w_{x\rightarrow y}$ indicates the weight of the directed link
from $x$ to $y$, which can also be interpreted as the coupling
strength. A simple illustration is shown in Fig.~\ref{fig1}. The
initial network as shown in Fig.~\ref{fig1}(a) is constituted of
seven nodes and eight directed links. Assuming the initial links'
weights are all equal to unit. After merging the three nodes $a$,
$b$ and $c$, a new node $m$ is generated, and according to
Eq.~\ref{weight} the links weights in the reduced network are drawn
as in Fig.~\ref{fig1}(b). Indeed, since the three nodes in total
receive three in-links from node $d$, while two from node $e$, the
weights of $m$'s two in-links are respectively $w_{d\rightarrow
m}=3/3=1$ and $w_{e\rightarrow m}=2/3$. For $m$'s out-links, since
the three nodes have two out-links to node $f$ and one to $g$, the
weight are respectively $w_{m\rightarrow f}=1+1=2$ and
$w_{m\rightarrow g}=1$.

Under the framework of master stability analysis, the
synchronizability of an undirected network can be quantified by the
ratio between the largest and the smallest non-zero eigenvalues of
the Laplacian matrix of this network, namely
$R=\lambda_{N}/\lambda_{2}$, where $\lambda_N$ and $\lambda_2$ are
respectively the largest and the smallest non-zero eigenvalues of
the Laplacian matrix~\cite{PRL054101,PRL2109,PRE5080}. In directed
networks, since the Laplacian matrix, defined as
$L_{ij}=k_{i}^{in}\delta_{ij}-a_{ij}$, is asymmetric with zero
rowsum, it has complex eigenvalues. In order to achieve the
synchronization condition, every eigenvalue is entirely contained in
the region of negative Lyapunov exponent for the particular master
stability function. Once the stability zone is bounded and the
imaginary part of complex eigenvalue is small enough, the network
synchronizability can be approximately measured by the real part of
eigenvalue ratio $R=\lambda_{N}^{r}/\lambda_{2}^{r}$, where
$\lambda_{N}^{r}$ and $\lambda_{2}^{r}$ are respectively the largest
and the second smallest real parts of eigenvalues~\cite{PRL138701,PRL228702,EPJB217}. Generally speaking, the stronger
the synchronizability, the smaller the ratio $R$. Note that an
accurate index for measuring the synchronizability in directed
networks has not yet been proposed and asks for further studies.
Here, we use the approximate index
$R=\lambda_{N}^{r}/\lambda_{2}^{r}$ as an indicator to see whether
the synchronizability of a directed network changes after coarse
graining. Usually $\lambda_{N}$ is proportional to the largest
degree $k_{\text{max}}$ (i.e., largest node's strength in weighted
network) of the network and $\lambda_{2}$ corresponds to the lowest
degree $k_{\text{min}}$ (i.e., lowest node's strength in weighted
network)~\cite{PR93}. Therefore, keeping the $k_{\text{max}}$ and
$k_{\text{min}}$ unchanged can effectively help to maintain the
synchronizability after coarse graining. Thus, in the coarse
graining process, the nodes with largest and smallest in-degrees can
only be merged if the $k_{\text{max}}$ and $k_{\text{min}}$ of the
coarse-grained network are respectively equal to that of the initial
network. Otherwise, we will randomly selected two nodes, one with
largest in-degree and the other with the smallest in-degree, before
group partition. Then the rest $N-2$ nodes will be classified into
$K-2$ groups according to k-means clustering. Note that, unless
stated otherwise, $k$ always refers to the in-degree. In appendix, we further discuss the effect of the constraint of keeping $k_{\text{max}}$ and $k_{\text{min}}$ on coarse graining results. It shows that the eigenvalue ratio $R$ is sensitive to $k_{\text{max}}$ and $k_{\text{min}}$, while the order parameter of Kuramoto model does not.

Finally, for the aspect of computational complexity, the k-means clustering algorithm is of $O(N^{2})$, and the time complexity of calculating the impact-vector $F$ is $O(N^{3})$. Likewise, the calculation of eigenvectors in SCG method also takes $O(N^{3})$. However, with the development of computing techniques, lots of fast calculation algorithms can help to reduce the computational complexity and make our method be able to deal with large networks. For example, the computational complexity of Katz index (i.e., the case for $\beta<1$) can be reduced to $O(N+M)$ where $M$ is the number of edges in the network~\cite{CMOT275}.

\section{Results}
\subsection{Coarse graining on modeled networks}

We apply the Path-based Coarse Graining (PCG) approach to four
kinds of directed networks: (i) Directed tree network. A tree with
$N$ nodes and $L$ layers is generated starting from a directed train
with length $L$, in which each node represents a layer. Then rest
$N-L$ nodes are added one by one. Each new added node is connected
by a directed link starting from one of its ancestors which are not
located in the layer $L$. (ii) A variant of Barab\'{a}si-Albert
networks~\cite{BA}: Directed BA network. An acyclic directed BA
network is generated by using the mechanism for undirected BA
network and assuming the link direction can only from older node to
younger node. (iii) A variant of Watts-Strogatz networks~\cite{SW}:
Directed WS network. The model starts from a completely regular
network with identical degree and clockwise links. Each link will be
rewired with two randomly selected nodes with probability $q$ ($\in
(0,1)$). (iv) A variant of Erd\"{o}s-R\'{e}nyi networks~\cite{ER}:
Directed ER random network. The directed ER random networks can be
generated by setting $q=1$.

\begin{figure}
  \center
  \includegraphics[width=4.2cm]{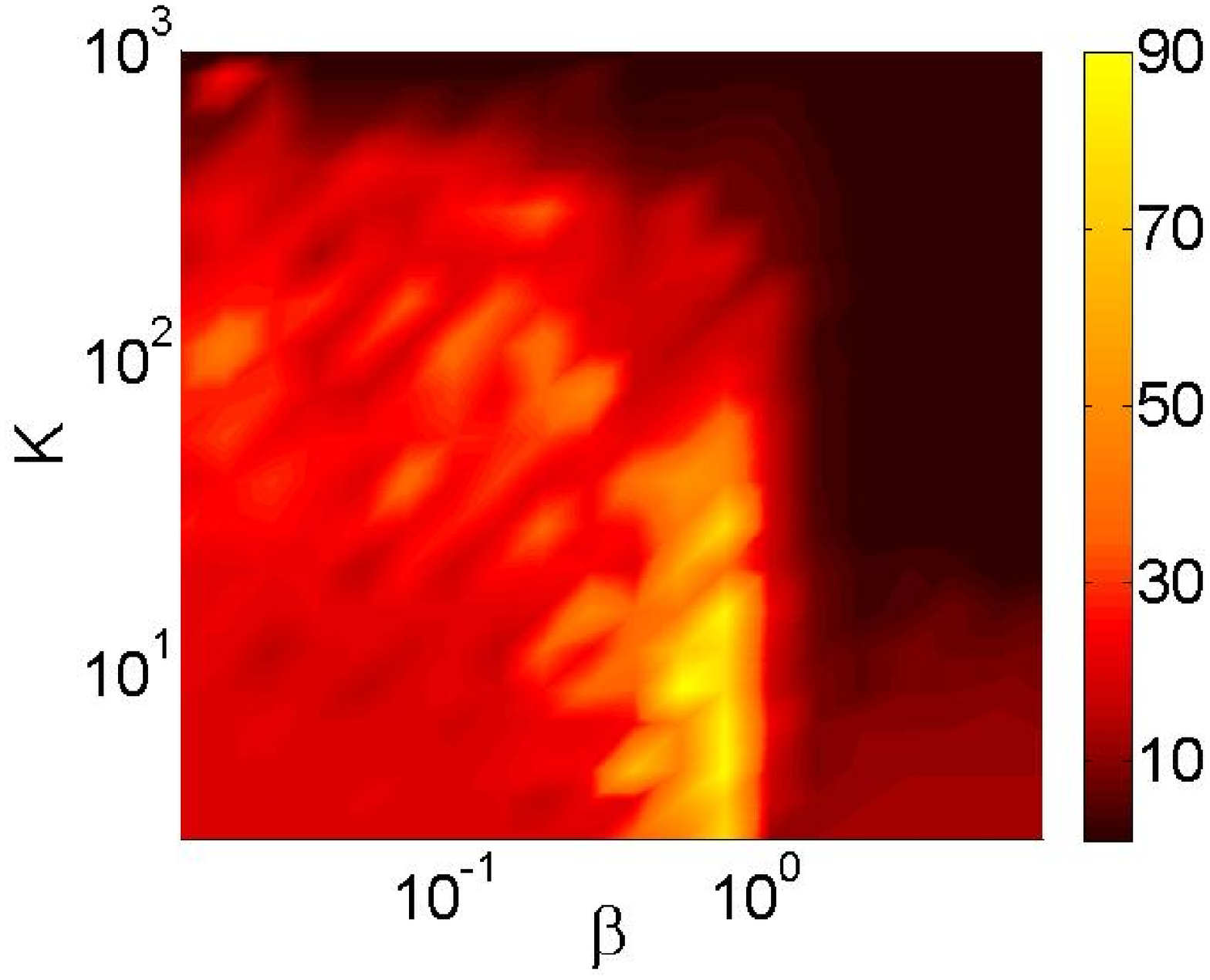}
  \includegraphics[width=4.2cm]{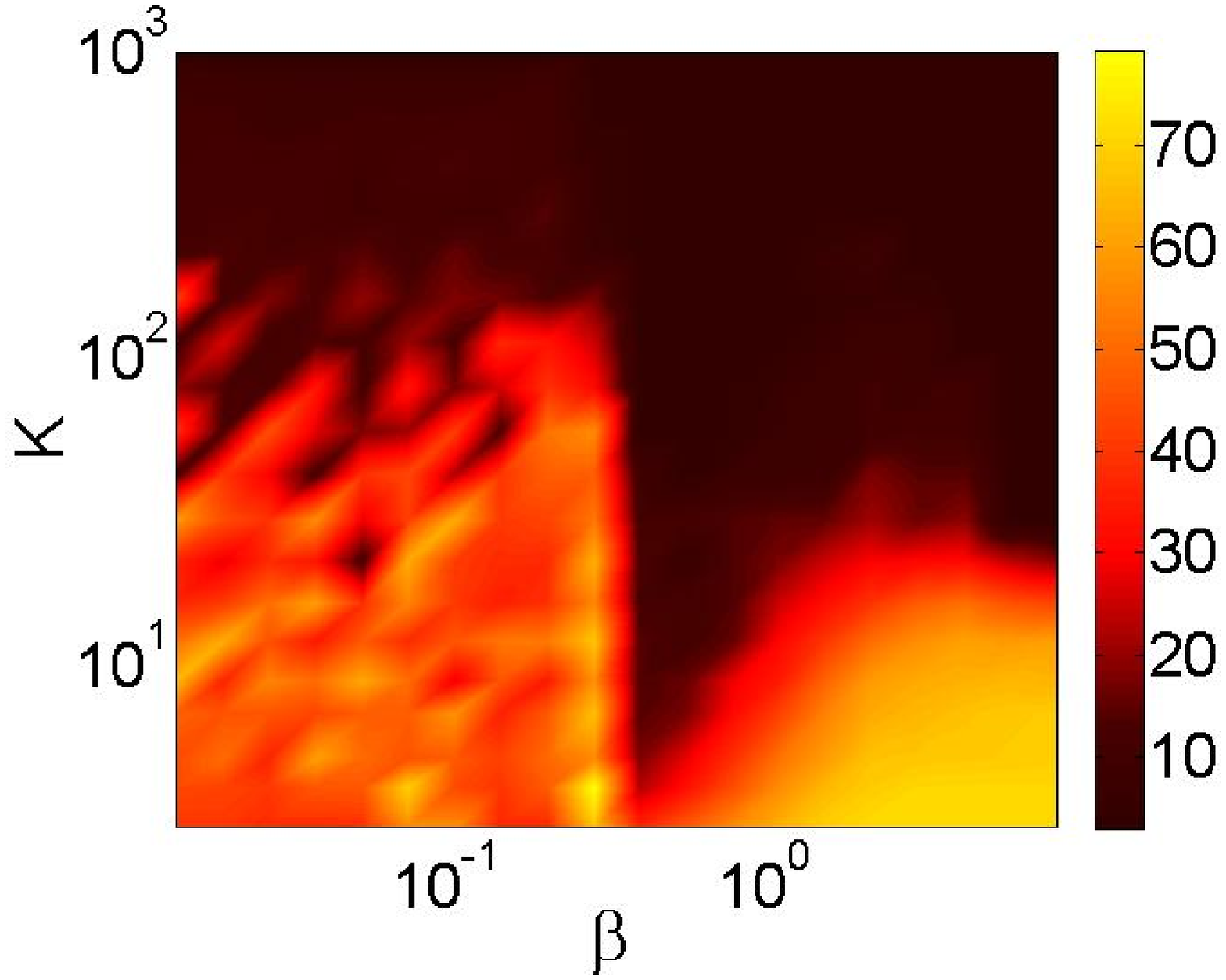}
  \mbox{(a)Tree network\hspace{2cm}(b)BA network}
  \includegraphics[width=4.2cm]{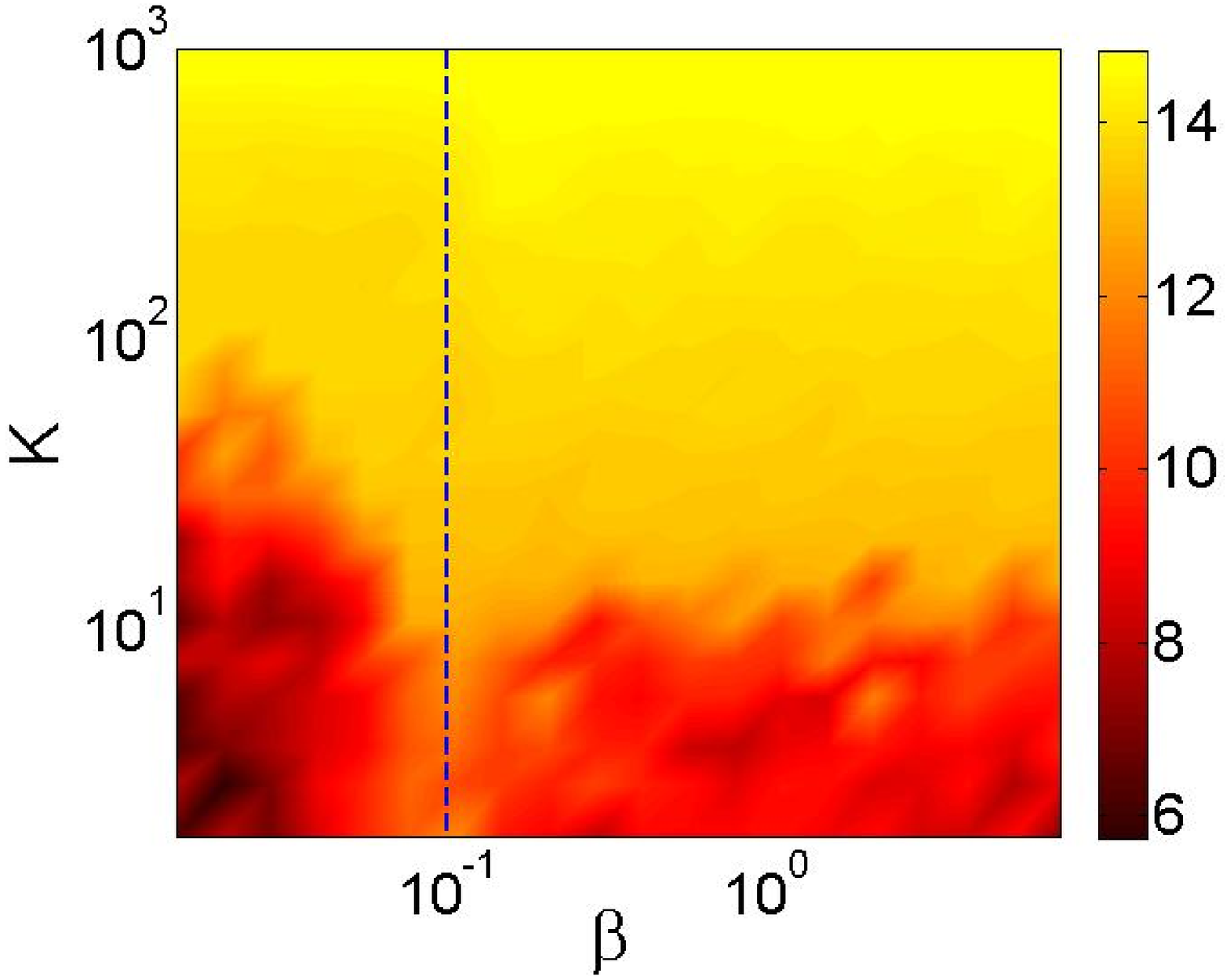}
  \includegraphics[width=4.2cm]{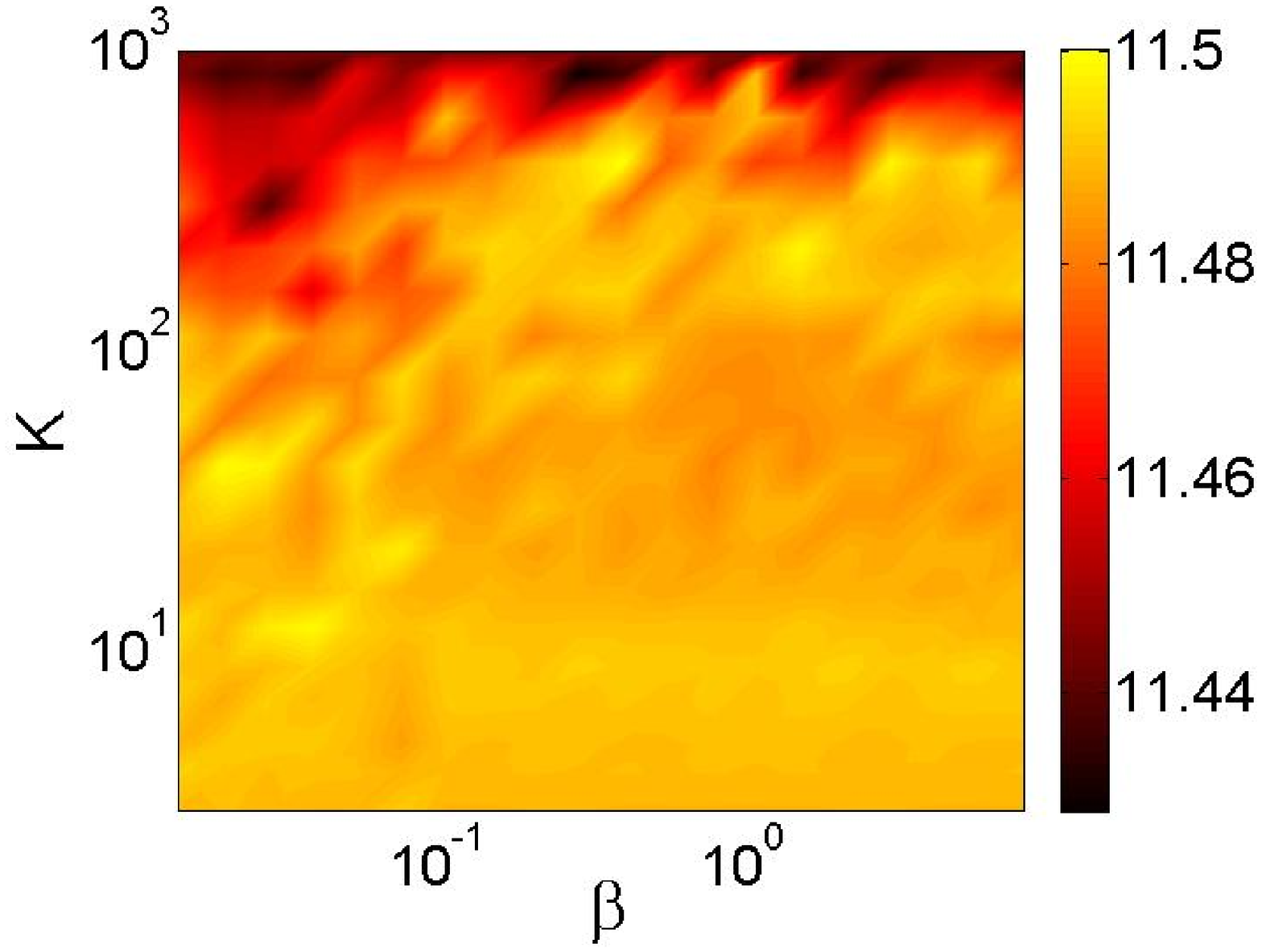}
  \mbox{(c)WS network\hspace{2cm}(d)ER network}
\caption{(Color online) The synchronizability $R$ in the
($\beta$,$K$) plane for (a) directed tree networks ($N=1000$,
$L=20$), (b) directed BA network ($N=1000$, $\bar{k}=3$), (c)
directed WS network ($N=1000$, $\bar{k}=10$, $q=0.1$) and (d)
directed ER network ($N=1000$, $\bar{k}=10$).}\label{fig5}
\end{figure}

Firstly, we investigate the performance of PCG on above four kinds
of networks. The
synchronizability of the coarse-grained network $R$ in the
$(\beta,K)$ plane is shown in Fig.~\ref{fig5}, where $K$ is the size of the coarse-grained network. Interestingly, we
find that in tree network and acyclic BA network larger $\beta$ in
average provides better results than smaller $\beta$. Especially, in
tree network there is an obvious line at $\beta \approx1$. In the BA
network, with $\beta>1$ the coarse-grained network can keep the
synchronizability exactly the same as the initial network. In the
cyclic WS network the $\beta$ that best preserves the
synchronizability is around 0.1. It seems that networks with more
loops tend to obtain better coarse graining with smaller $\beta$
(see subsection C for detailed discussion of the relationship
between the optimal parameter $\beta^*$ and the number of loops in
network). The result in Fig.~\ref{fig5}(d) shows that the
synchronizability of the coarse-grained ER network is not sensitive
to $\beta$ regardless of $K$, since the total fluctuation is smaller
than 0.07.

We compare the PCG method with other two methods, namely Random
Coarse Graining (RCG) and Spectral Coarse Graining (SCG)~\cite{PRL174104}. In RCG, the $N$ elements of each node's vector are
randomly selected in the range of (0,1). Then the nodes will be
classified into $K$ groups by using k-means clustering. In directed
networks, the egeinvalues and egeinvectors of their laplacian
matrixes have complex values. When we apply SCG to
directed networks, we consider only the real parts of the values in this paper. In practice, we define $I$ equally distributed intervals between the
maximum and minimum of $p_{2}^{r}$ ($p_{N}^{r}$), where $p_2^r$ and
$p_N^r$ are the egeinvectors corresponding to the second smallest
and the largest real-part-egeinvalues of the laplacian matrix,
respectively. The nodes whose eigenvector components in $p_2^r$
($p_N^r$) fall in the same interval are merged. Specifically, if the
elements in both $p_2^r$ and $p_N^r$ are identical, we will randomly
divide the nodes into $K$ groups. Actually, the relation between $I$
and $K$ strongly depends on the network structure (i.e., the
distribution of the elements in $p_2^r$ and $p_N^r$). For instance,
considering the initial WS and ER network shown in Fig.~\ref{fig5},
when $I=800$ the size of the coarse-grained WS network is 951, while
the reduced ER network only contains 281 clusters. Note that there are many potential ways to apply the SCG method to directed networks making use of the imaginary part of the elements in eigenvectors. For example, the vectors $p_{2}$ ($p_{N}$) can be generated by combining the real part and the imaginary part (such as $\sqrt{(p_{2}^{r})^{2}+(p_{2}^{i})^{2}}$, $p_{2}^{r}+p_{2}^{i}$, $p_{2}^{r}p_{2}^{i}$, et al.). Another way is grouping the nodes according to four vectors, namely $(p_{2}^{r},p_{2}^{i},p_{N}^{r},p_{N}^{i})$. If the imaginary parts are appropriately considered, the performance of SCG can be improved. However, how to find the right way to make use of the imaginary parts is a tough problem and inappropriately involving the imaginary parts in SCG method may lead to even worse results.

\begin{figure}
  \center
  \includegraphics[width=9.2cm]{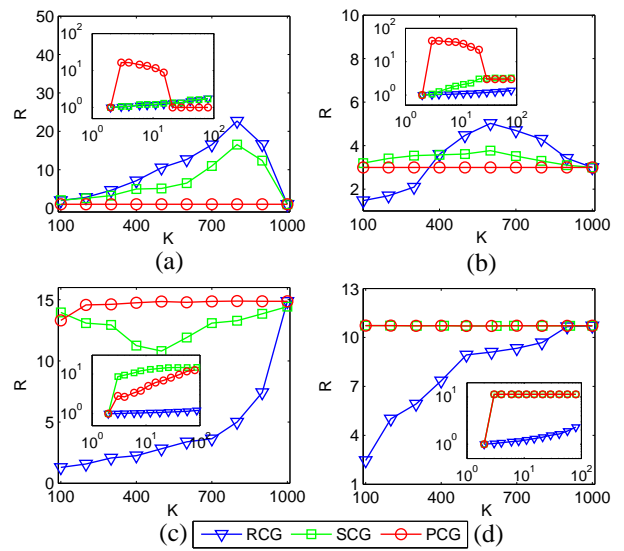}
\caption{(Color online) The evolution of the ratio
$R=\lambda_{N}^{r}/\lambda_{2}^{r}$ as a function of the size of the
coarse-grained network $K$. The initial networks are the same to the
ones in Fig.~\ref{fig5}. We use the typical parameter $\beta=5$ for
(a) directed tree networks and (b) directed BA networks, and
$\beta=0.1$ for (c) directed WS networks and (d) directed ER
networks. The results for RCG and PCG are obtained by averaging over
$100$ independent network realizations. Insets show the results for
$K\in[2,100]$. }\label{fig2}
\end{figure}

Figure~\ref{fig2} shows how the indicator $R$ changes with $K$ on
the above four kinds of directed networks with typical $\beta$.
Overall speaking, PCG outperforms SCG, and RCG performs worst. As
shown in Fig.~\ref{fig2}(a) (b) and (c), with SCG and RCG, the
synchronizability changes even only a few nodes have been merged,
while the PCG displays a large stable range. In a directed tree
network, most of the elements in eigenvectors corresponding to the
smallest and the largest eigenvalues are identical. Thus it is
impossible to distinguish the role of nodes by the analysis on the
eigenvectors as suggested in Ref.~\cite{PRL174104}. In the tree
networks, the most effective coarse graining strategy is to merge
the nodes in the same layer. PCG can indeed well identify the nodes
in different layers by using a larger parameter $\beta$ ($>1$). In
this sense, PCG is very effective in acyclic directed tree network.
From Fig.~\ref{fig2}(a), one can see that when $K>L$, PCG can keep
the synchronizability exactly the same as the initial network. When
$K=L$, the tree will be reduced to a train with length $L$, namely
all the nodes in the same layer are merged. When $K<L$, there exist
a suddenly jump of $R$, see inset of Fig.~\ref{fig2}(a). This is
caused by merging the nodes in different layers and thus leading to
a smaller $k_{\text{min}}$ according to the weighting strategy in
Eq.~\ref{weight}. In this case, if we artificially set
$k_{\text{min}}$ of the reduced network equal to that of the initial
network, the synchronizability can be well preserved (exactly equal
to 1). Similar phenomenon exists in acyclic BA network where the
hierarchical structure is clear.

It has been demonstrated that the synchronizability of the directed
BA network with average in-degree $\bar{k}=3$ is exactly 3~\cite{PRL138701}. Figure~\ref{fig2}(b) shows that PCG with parameter
$\beta=5$ can guarantee $R=3$ by keeping the network acyclic and
$k_{\text{max}}$ and $k_{\text{min}}$ unchanged, even the network is
reduced to 30 clusters (i.e., $K=30$). When $K<30$, merging may
generate some loops and decrease $k_{\text{min}}$, and thus lead to
a sharp increase of $R$. It can not be perfectly avoided by
artificially keeping $k_{\text{min}}$ as what we did in the tree
network, instead $R$ can effectively reduce to around 3, since here
the loops also play a role. On the contrary, the SCG method may
induce loops even merging a few nodes (i.e., for a larger $K$). For
example, when $K=600$, the synchronizability of the reduced SCG
network is $R=3.77$, while synchronizability of the reduced PCG
network is exactly equal to 3.

In the networks with cycles including directed WS networks and
directed ER networks, there are no clear hierarchical structures,
thus the local information (i.e., short paths) plays more important
role to quantify the node's impact during the coarse graining
process, and thus a relative small $\beta$ is required. Here we use
$\beta=0.1$. The results show that the PCG method performs as well
as SCG method in directed ER networks while much better than SCG and
RCG in directed WS networks.

\begin{figure}
  \center
  \includegraphics[width=4cm]{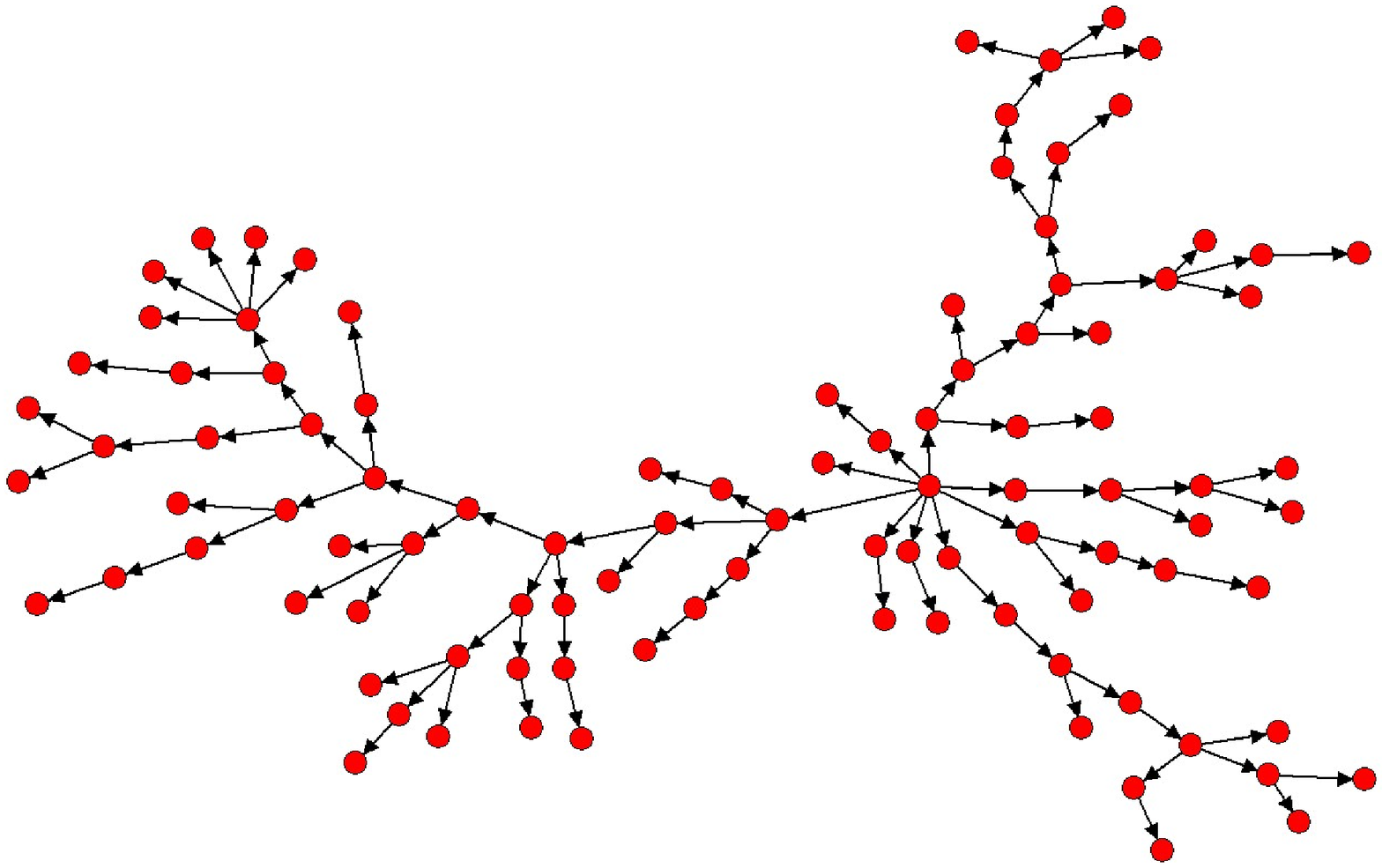}
  \includegraphics[width=4cm]{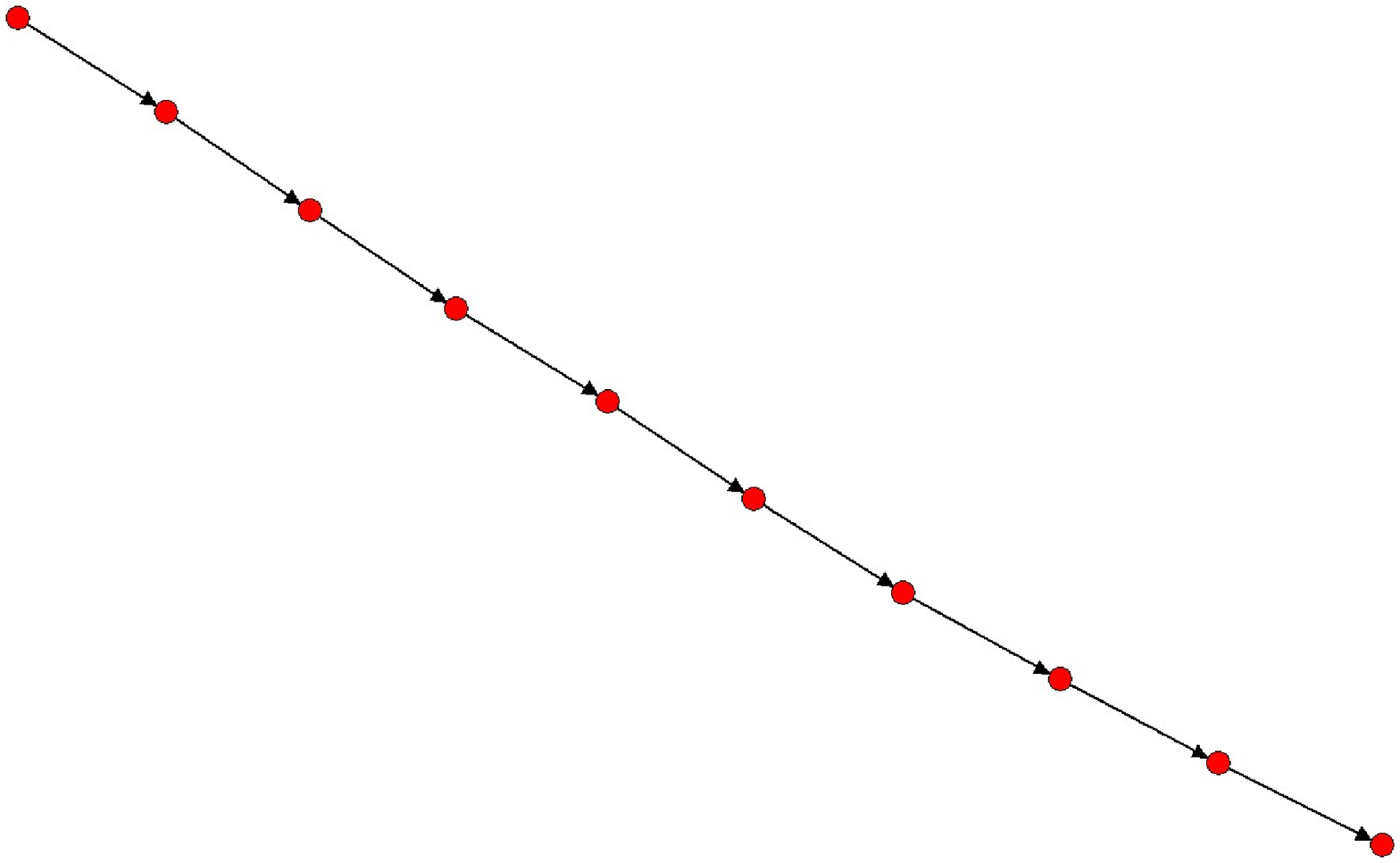}\\
  \mbox{(a)\hspace{4cm}(b)}
  \includegraphics[width=4cm]{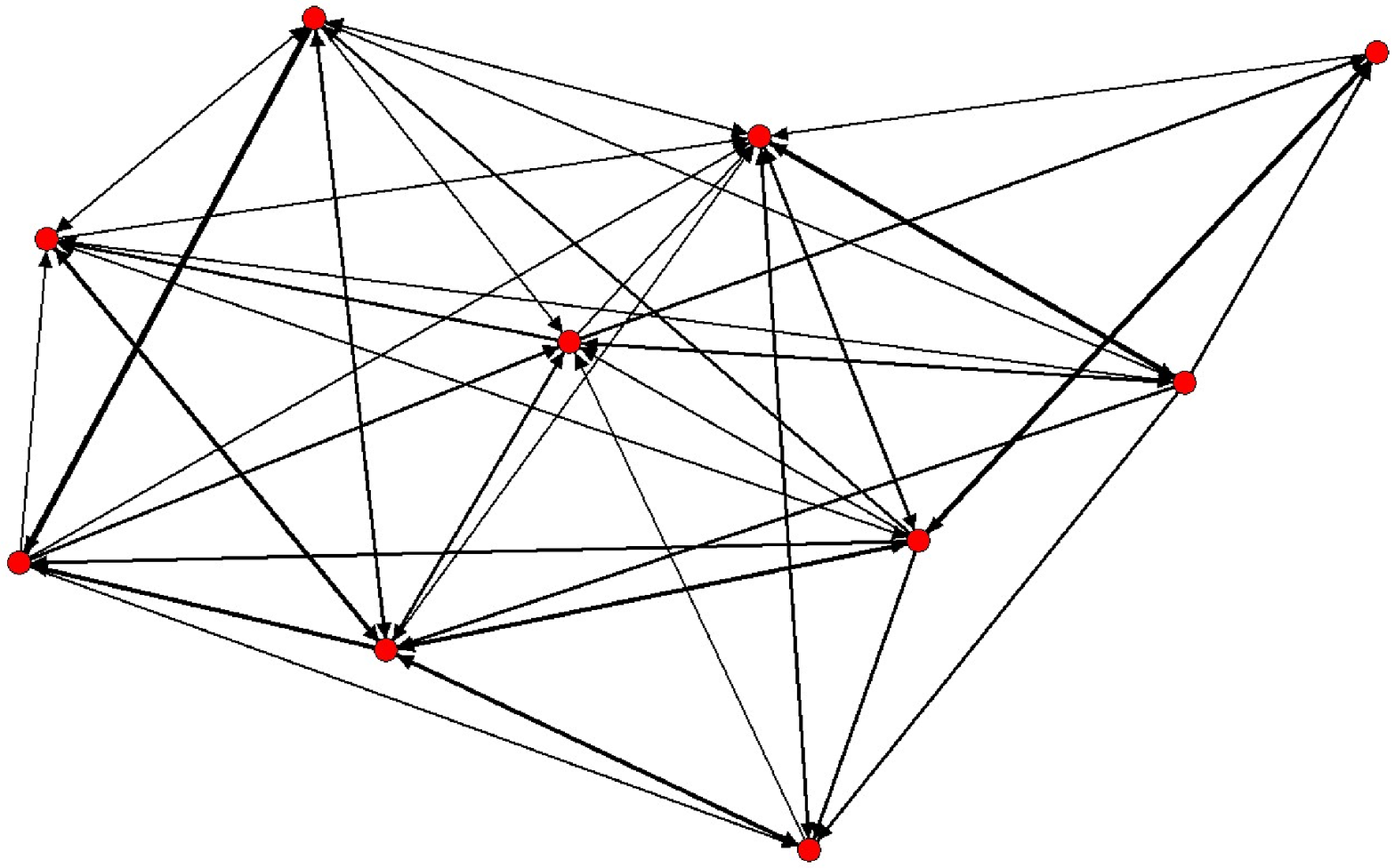}
  \includegraphics[width=4cm]{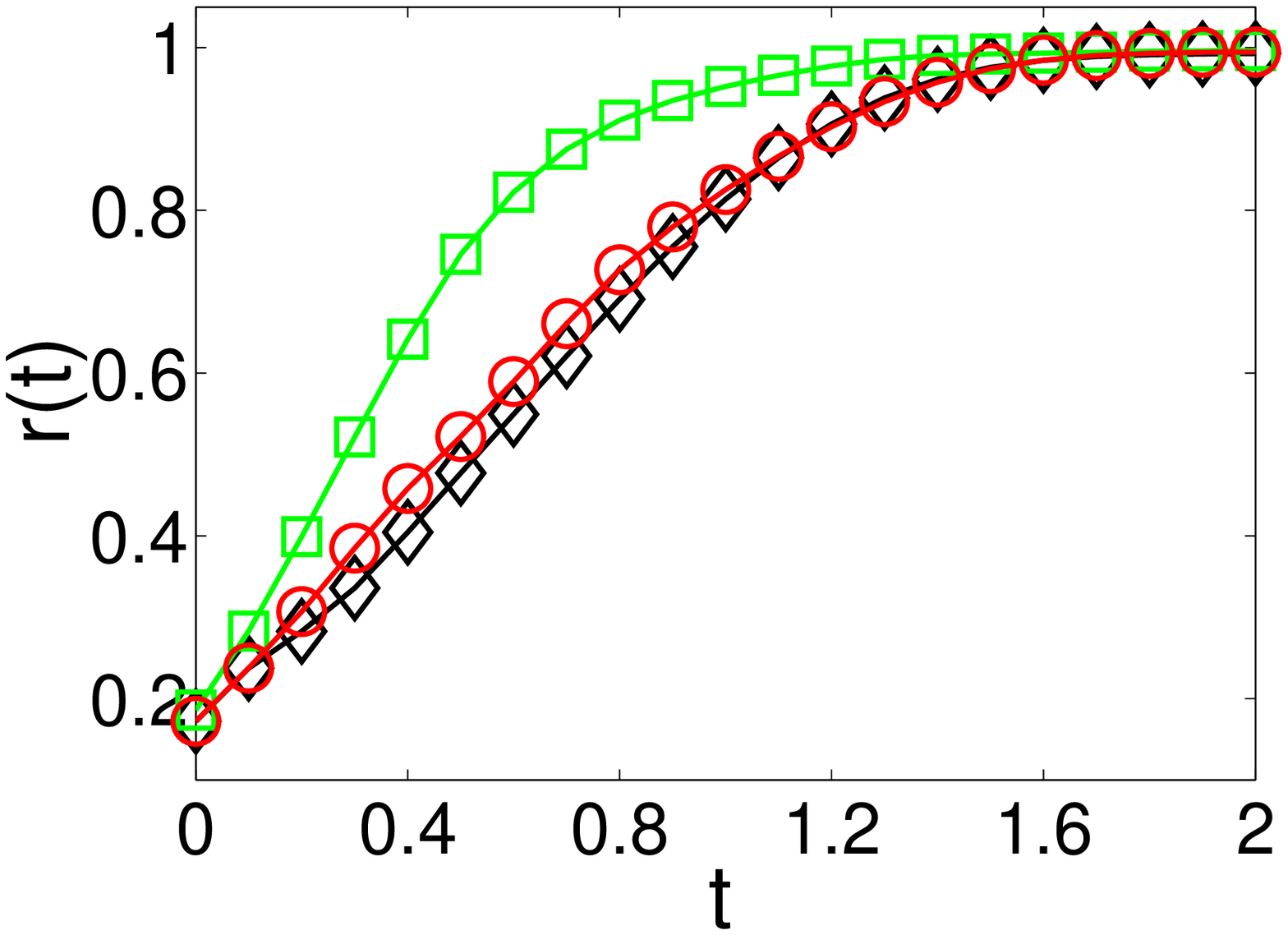}\\
  \mbox{(c)\hspace{4cm}(d)}
  \includegraphics[width=4cm]{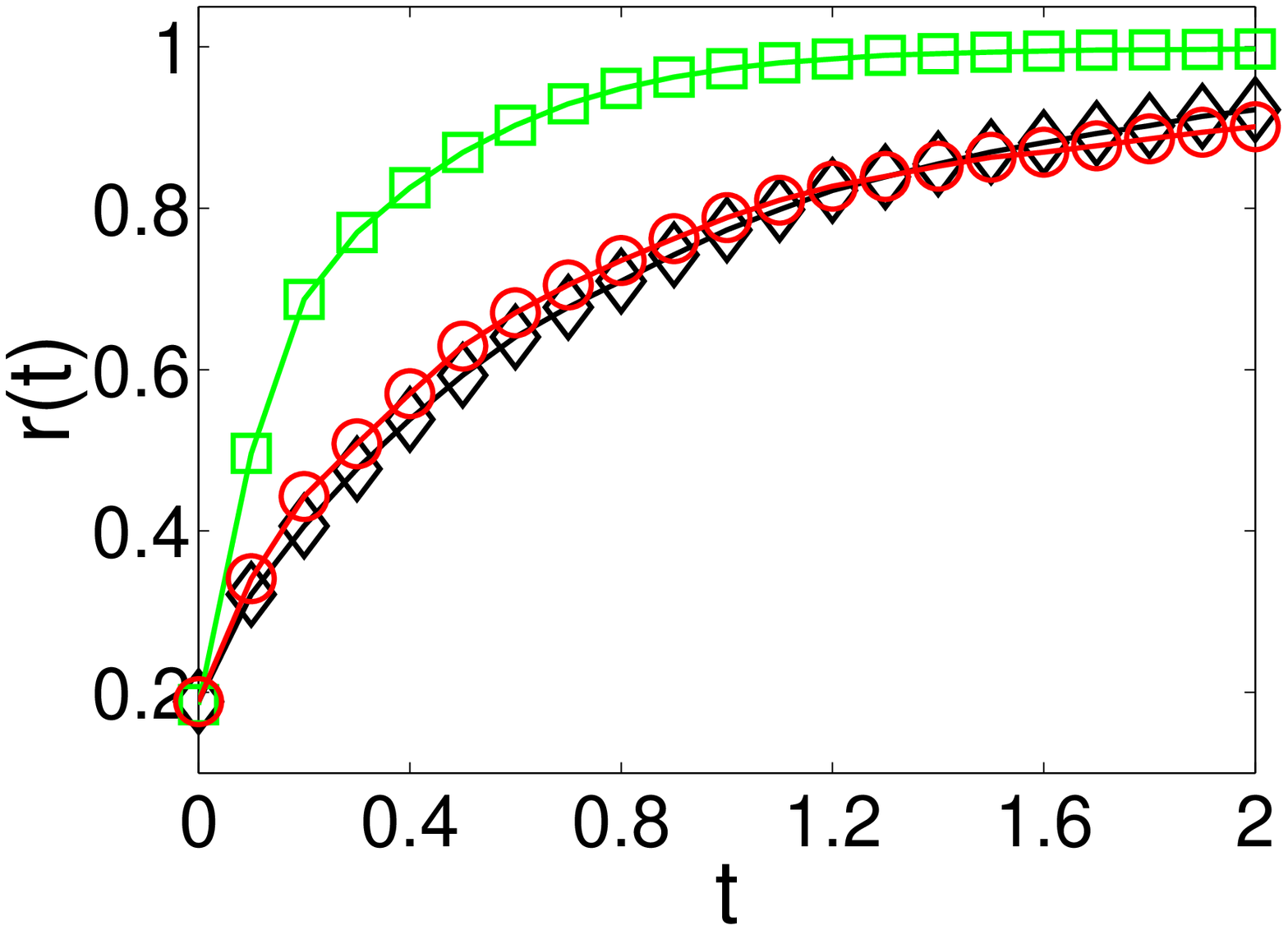}
  \includegraphics[width=4cm]{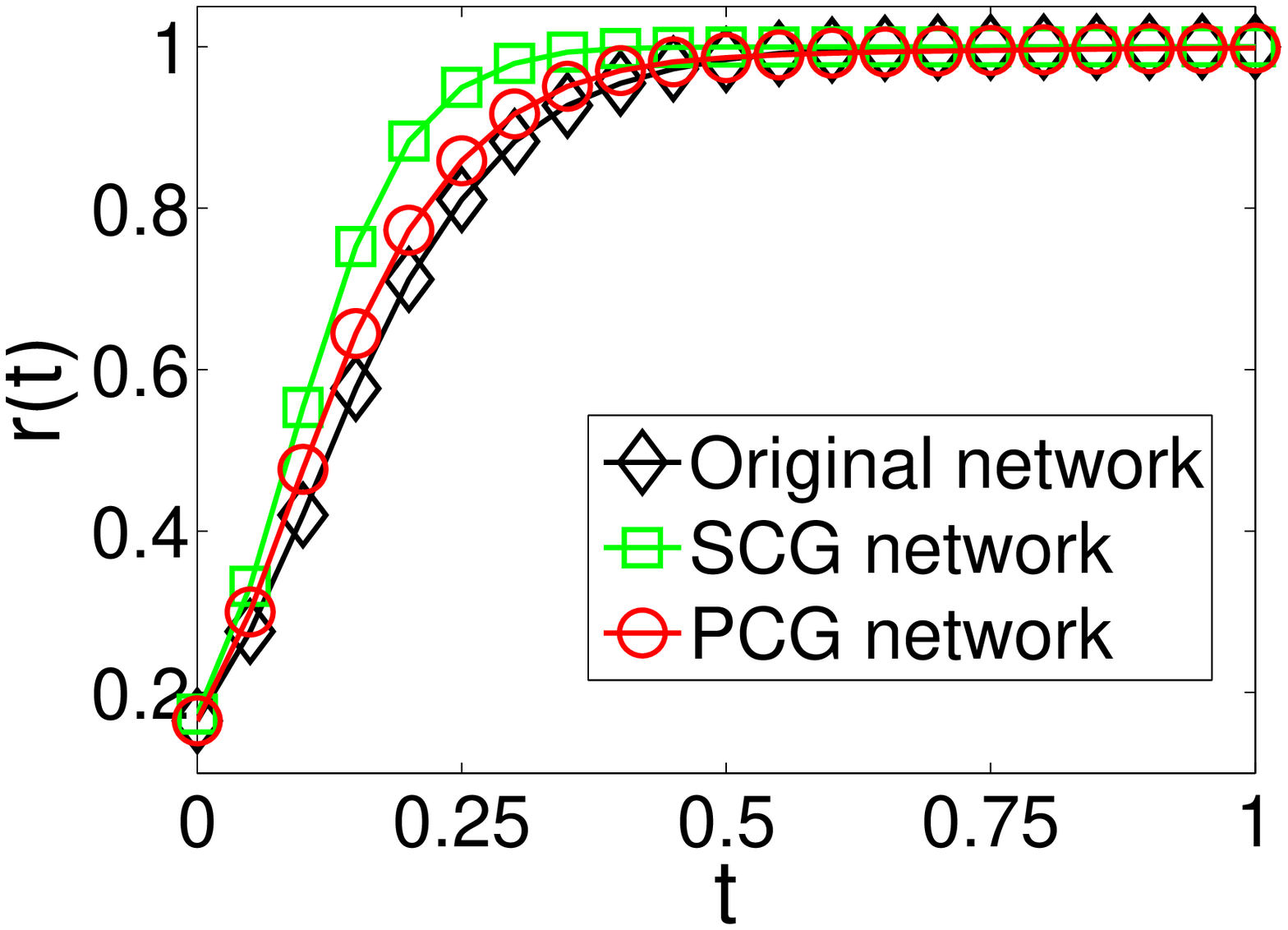}\\
  \mbox{(e)\hspace{4cm}(f)}
\caption{(Color online) Given a specific directed tree network with
$N=100$ nodes and $L=10$ layers as shown in (a), the coarse-grained
networks through PCG and SCG are respectively presented in (b) and
(c) which are constituted of $K=10$ clusters. Figure (d) shows the
performance of Kuramoto model on these three networks, namely (a)
original network, (b) PCG network and (c) SCG network. Figures (e)
and (f) show respectively the results of WS network (with $N=100$,
$\bar{k}=4$, $q=0.1$) and BA network (with $N=100$ and $\bar{k}=3$).
Their coarse-grained networks all contain $K=25$ clusters. The
coupling strength is $\sigma=10$ and $w_i$ is randomly selected in
the range of $(-0.5,0.5)$. Initially, $\theta_i$ is randomly chosen
in $(-\pi,\pi)$.}\label{fig3}
\end{figure}

In addition, we point out that grouping the nodes aiming at preserving the dynamics cannot maintain the topological properties at the same time, although the grouping is according to the topological similarity. Generally, the average degree of the coarse-grained network is larger than that of the initial network. For comparison, we generate a group of modeled networks which have the same topological properties as the initial network and same size as the coarse-grained network. It is shown that the average number of reachable nodes and loop number of the coarse-grained networks are always higher than that of the modeled networks, while the average shortest distance of the coarse-grained networks is always smaller than the modeled networks. Moreover, the coarse graining procedure may change the degree distribution of the initial networks. However, the topological properties of the PCG networks are relatively closer to the initial networks than the SCG networks especially in the acyclic networks (not so obvious in directed networks with cycles). For example, the PCG method can prevent the producing of loops and keep the coarse-grained networks still partial reachable.

\subsection{Kuramoto model on coarse-grained networks}

Since the Laplacian matrixes for directed networks are asymmetric,
the egeinvalues $\lambda_{2}$ and $\lambda_{N}$ are complex. In this
case, the indicator $R$ can not exactly represent the
synchronizability of a network. Hence, we further test our method
with the Kuramoto model~\cite{Kuramoto1975,Acebron2005}, which is a
classical model to investigate the phase synchronization phenomena.
The coupled Kuramoto model in the network can be written as
\begin{equation}
\dot{\theta_{i}}=\omega_{i}+\sigma\sum\limits_{j=1}\limits^{N}
A_{ij}sin(\theta_{j}-\theta_{i}),\quad \quad i=1,2.....,N
\end{equation}
where $\omega_{i}$ and $\theta_{i}$ are the natural frequency and
the phase of oscillator $i$ respectively, and $A$ is the adjacency
matrix. The collective dynamics of the whole population is measured
by the macroscopic complex order parameter,
\begin{equation}
r(t)e^{i\phi(t)}=\frac{1}{N}\sum\limits_{j=1}\limits^{N}e^{i\theta_{j}(t)},
\end{equation}
where the modulus $r(t)$ ($\in[0,1]$) measures the phase coherence
of the population and $\phi(t)$ is the average phase. $r(t)\simeq1$
and $r(t)\simeq0$ describe the limits in which all oscillators are
respectively phase locked and moving incoherently. By studying the
behavior of the order parameter $r(t)$, we are able to investigate
whether the topology of the coarse-grained network is representative
of the initial one. The initial network is a tree network as
shown in Fig.~\ref{fig3}(a), which contains 100 nodes and has 10
layers. After the PCG procedure, we obtain a train-like network with
depth equal to 10, see Fig.~\ref{fig3}(b). With the SCG method, a
cyclic network will be generated as shown in Fig.~\ref{fig3}(c). The
result of how the order parameter $r(t)$ of Kuramoto model performs
in these three networks is shown in Fig.\ref{fig3}(d). It is obvious
that $r(t)$ of the PCG network converges with almost the same speed
as the initial one, while in the SCG network it converges faster. Moreover, the results of directed WS network and BA network
are respectively shown in Fig.~\ref{fig3}(e) and (f). Their
coarse-grained networks all contain 25 clusters. It is clearly that
the PCG method can preserve the synchronizability more effectively
than SCG.

\subsection{The optimal parameter $\beta^*$ for different networks}

\begin{figure}
  \center
  \includegraphics[width=4.2cm]{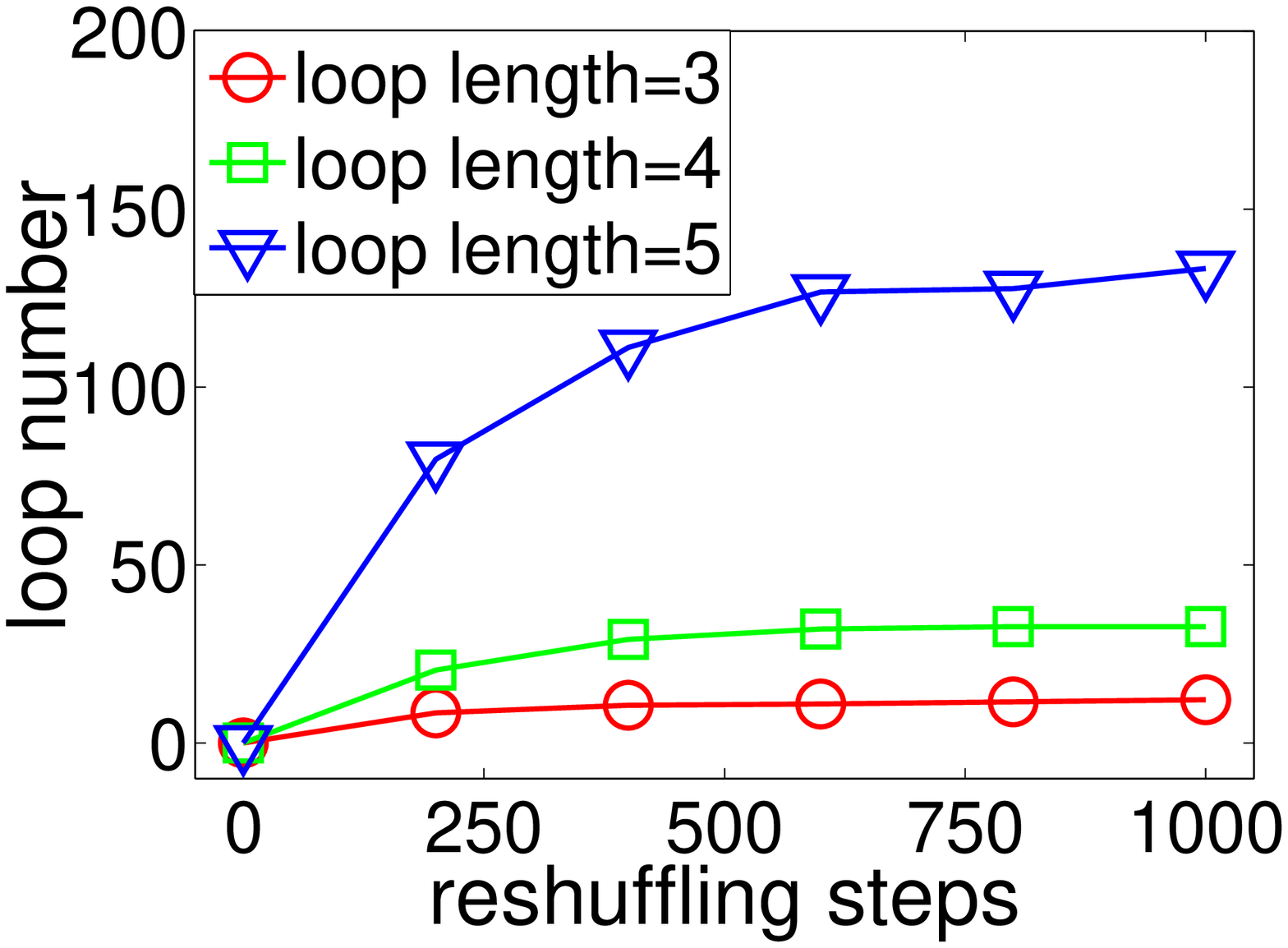}
  \includegraphics[width=4.2cm]{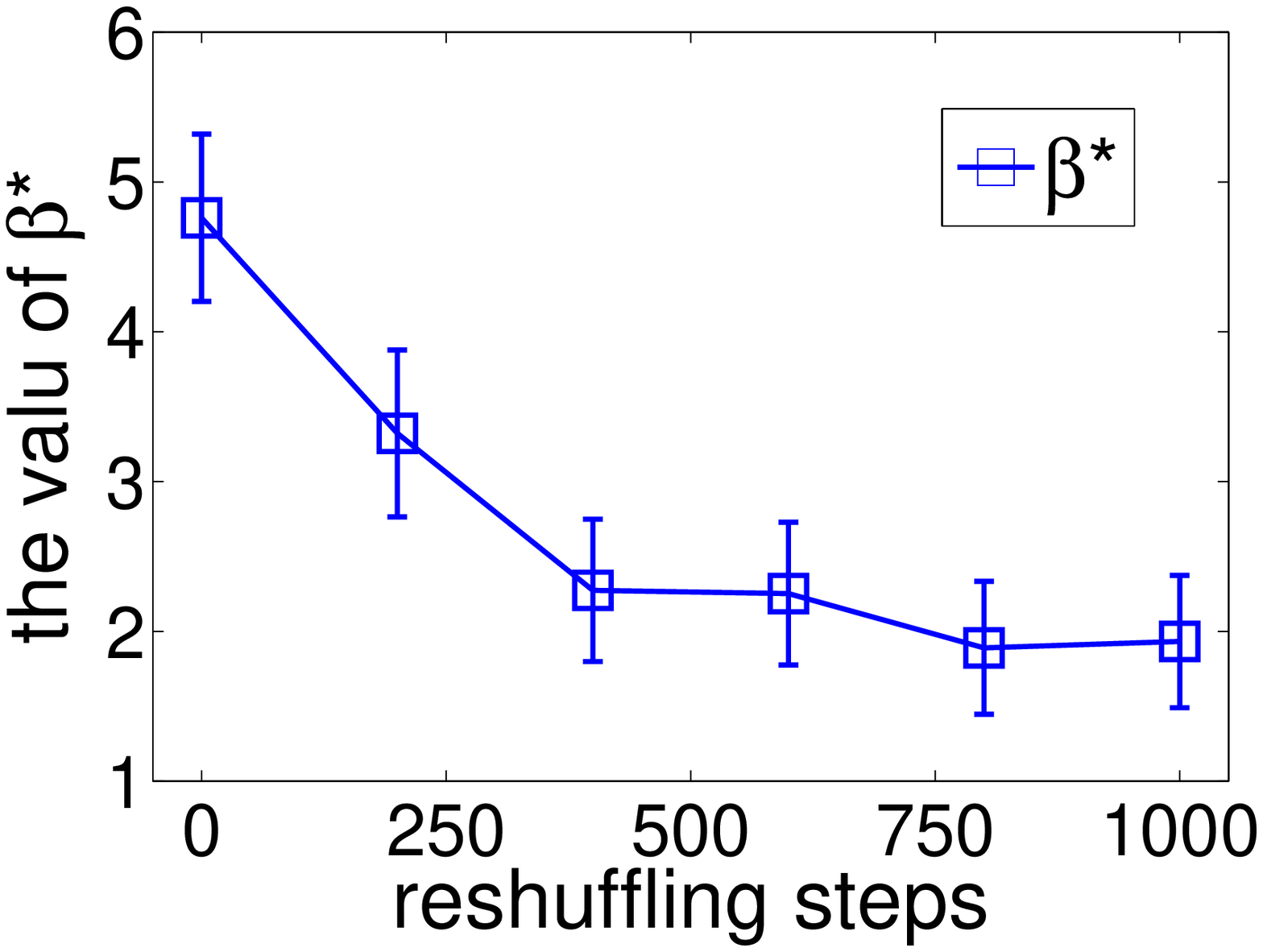}
  \mbox{(a)\hspace{4cm}(b)}
\caption{(Color online) (a) The dependence of the number of loops
with different length on the reshuffling steps in directed BA
network ($N=100$, $\bar{k}=5$). (b) The $\beta^{*}$ as function of
the reshuffling steps. Each point is obtained by averaging over
$100$ independent network realizations.}\label{fig4}
\end{figure}

In different networks, the optimal parameters $\beta^*$
corresponding to the best performance on coarse graining are
different. Empirically, the $\beta^*$ of acyclic networks is larger
than that of those containing loops. To investigate whether
$\beta^{*}$ is affected by the cycles in networks, we carry out an
experiment based on directed BA networks, on which loops are
generated by reshuffling some links. Specifically, we randomly
select two directed links from the network, for example, one is from
node $A$ to $B$ and the other is from node $C$ to $D$. Then we
rewired these two links as $A$ to $D$ and $C$ to $B$. In this way,
the degree of these nodes will not be changed during the reshuffling
procedure. In average, reshuffling more links leads to more loops,
see an example in Fig.~\ref{fig4}(a) where the numbers of loops with
length 3, 4 and 5 all increase with the increasing of reshuffling
steps. Now, we would like to find the optimal parameters for the
reshuffled networks. For a given network, $\beta^*$ might be
different with different $K$ as we have shown in Fig.~\ref{fig5}.
However, in practice, checking the optimal parameter for different
$K$ in advance is sometimes impossible. Thus, we here ignore the
relationship between $\beta$ and $K$, and consider the general
performances of one parameter on the coarse-grained networks with
the possible sizes we concerned. The $\beta^{*}$ is thus
corresponding to the $\beta$ that yields the minimum
synchronizability difference between the coarse-grained networks and
the initial networks, which can be mathematically expressed by:
\begin{equation}
d=\sum\limits_{K=n}\limits^{N}|R_{K}-R_{0}|
\end{equation}
where $R_{K}$ is the synchronizability of the coarse-grained network
with $K$ nodes, $R_{0}$ is the synchronizability of the initial
network and $n$ is the minimum size of the coarse-grained network
that we considered. Since too small $K$ may lead to dramatic change
of $R$, here we choose $n=10$ in the example shown in
Fig.~\ref{fig4}. We obtain $\beta^{*}$ subject to the minimum $d$.
The dependence of $\beta^{*}$ on the number of reshuffling steps is
shown in Fig.~\ref{fig4}(b). Instead of considering all possible
$\beta$ which is very time consuming, we test the parameter $\beta$
in the range of [0.01,10] with step 0.01, 0.1 and 1 respectively in
[0.01,0.1), [0.1,1) and [1,10]. It is clear that $\beta^{*}$
decreases with the increasing of reshuffling steps. Actually, if directed networks have obvious hierarchical structure and rare loops, PCG can perform better with a relatively large $\beta$ since it emphasizes on long path to detect the hierarchical structure. However, in directed networks with many loops, the hierarchical structure is not clear. As a path involved in loops can be regarded as an infinite long path, its effect on the impact-vector will be enormously amplified with a large $\beta$, and thus leading to noise when characterizing the dynamic role of a node. In this case, it is better to pay more attention to the impacts from local structure, namely emphasize the effects of short paths by using small $\beta$.

\section{Conclusion}

Coarse graining is an effective way to analyze and visualize large
networks. Many methods and models have been proposed to reduce the
size of the networks and preserve main properties such as degree
distribution, cluster coefficient, degree correlation, as well as
some dynamic behaviors such as random walks, synchronizability and
critical phenomena. However, most of these works take into account
the undirected networks, while the study on coarse graining of
directed networks lacks of attention. In this paper, we introduce a
Path-based Coarse Graining (PCG) method which assumes that two
nodes are structural-similar if they obtain the same impacts from
other nodes, and thus they are more likely to be merged during the
coarse graining process. The impacts that a node obtained from other
nodes are calculated via tracing the origin of impacts in directed
network. Specifically, the impact of node $x$ on node $y$ is defined
by summing over the collection of directed paths from $x$ to $y$
with exponential weights by length, which are controlled by a
parameter $\beta$. Larger $\beta$ indicates the long paths are more
important (i.e., assign more weights to the long paths). Numerical
analysis on four kinds of directed networks, including tree-like
networks and variants of Barab\'{a}si-Albert networks,
Watts-Strogatz networks and Erd\"{o}s-R\'{e}nyi networks, shows that
our method can effectively preserve the synchronizability during the
coarse graining process. This result is further demonstrated by the
Kuramoto model. In addition, we find that the long paths play more
important roles on the coarse graining in the tree-like networks,
while in the cyclic networks, the long paths that involve the loops
usually have negative effects on quantifying the impacts of one node
on the other nodes during the coarse graining process, and thus a
smaller parameter $\beta$ gives better performance.

Finally, we claim that the idea for merging nodes which receive the same impacts from the network is quite general for coarse graining directed networks. For example, for random walk, two nodes in a directed network having exactly the same upstream neighbors should be grouped together since their random walker probabilities come from the same sources. In this sense, coarse graining directed networks for other dynamics can be interesting extensions.

\section*{Acknowledgement}
We thank Yi-Cheng Zhang, Tao Zhou and Jie Ren for their helpful suggestions. This work is
supported by the Swiss National Science Foundation under Grant No.
(200020-132253).

\appendix

\section{PCG method without degree constraint}

\begin{figure}
  \center
  \includegraphics[width=9cm]{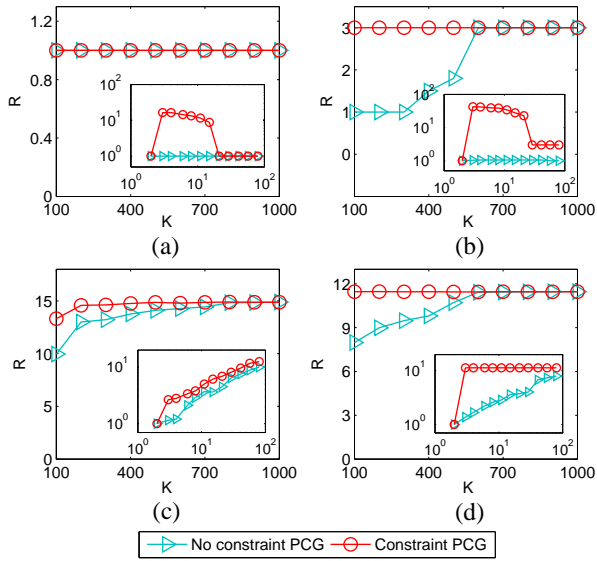}
\caption{(Color online) Comparison of the PCG with and without the constraint of $k_{\rm max}$ and $k_{\rm min}$. All the parameters in this figure are the same to the
ones in Fig.~\ref{fig5}.}\label{fig6}
\end{figure}

\begin{figure}
  \center
  \includegraphics[width=9cm]{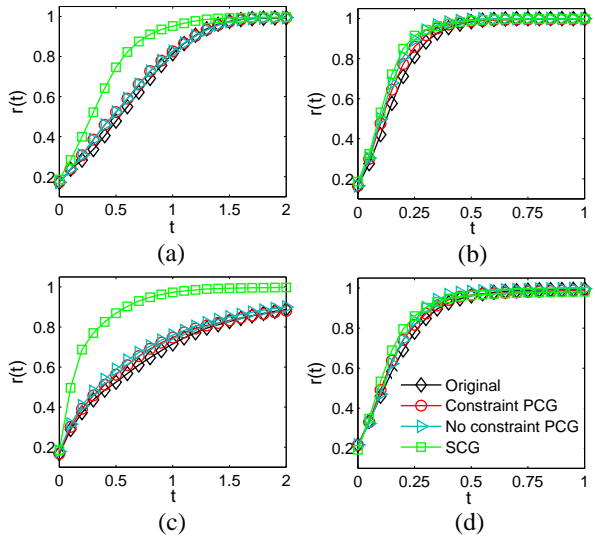}
\caption{(Color online) The order parameter of Kuramoto model on four networks. All the parameters in this figure are the same to the
ones in Fig.~\ref{fig3}.}\label{fig7}
\end{figure}

In the paper, we assumed that the nodes with largest and smallest in-degrees can
only be merged if the $k_{\text{max}}$ and $k_{\text{min}}$ of the
coarse-grained network are respectively equal to that of the initial
network. In order to investigate the effect of keeping the maximum and minimum in-degree on the coarse graining result, we remove the constraint of $k_{\rm max}$ and $k_{\rm min}$ in PCG and see the performance of the modified PCG method. As we mainly consider synchronization, the indicator $R$ is shown to be sensitive to the $k_{\rm max}$ and $k_{\rm min}$, see Fig.~\ref{fig6}. It is obvious that the PCG with constraint performs better than that without constraint. However, as shown in Fig.~\ref{fig7}, the order parameter $r(t)$ of the Kuramoto model does not show obvious differences. Moreover, we compared the RCG with and without the in-degree constraint. The result shows that the degree constraint cannot prominently improve the performance of RCG.

\end{document}